\documentclass{emulateapj}

\shorttitle{Formation of the outer stellar halo}
\shortauthors{Murante et al.}

\begin{document}


\title{ Assembly of the outer Galactic stellar halo \\
    in the hierarchical model}


\author{Giuseppe Murante}
\affil{Osservatorio di Torino,
    Strada Osservatorio 20, I-10025, Pino Torinese (TO), Italy}
\email{murante@oato.inaf.it}

\author{Eva Poglio}
\affil{Dipartimento di Fisica Generale ``Amedeo Avogadro'', Universit\`a degli
  Studi di Torino, Via P. Giuria 1, I-10125, Torino (Italy)}
\email{epoglio@studenti.ph.unito.it}

\author{Anna Curir}
\affil{Osservatorio di Torino,
    Strada Osservatorio 20, I-10025, Pino Torinese (TO), Italy}
\email{curir@oato.inaf.it}

\and

\author{\'Alvaro Villalobos}
\affil{I.N.A.F, Osservatorio di Trieste,
    Via Tiepolo 11, I- 34131, Trieste, Italy}
\email{villalobos@oats.inaf.it}


%\altaffiltext{1}{Visiting Astronomer, Cerro Tololo Inter-American Observatory.
%CTIO is operated by AURA, Inc.\ under contract to the National Science
%\altaffiltext{2}{Society of Fellows, Harvard University.}
%\altaffiltext{3}{present address: Center for Astrophysics,
%    60 Garden Street, Cambridge, MA 02138}
%\altaffiltext{4}{Visiting Programmer, Space Telescope Science Institute}
%\altaffiltext{5}{Patron, Alonso's Bar and Grill}


\begin{abstract}
We provide a set of numerical N-body simulations for studying the
formation of the outer Milky Ways's stellar halo through accretion
events.  After simulating minor mergers of prograde and retrograde
orbiting satellite halo with a Dark Matter main halo, we analyze the
signal left by satellite stars in the rotation velocity distribution.
The aim is to explore the orbital conditions where a retrograde signal
in the outer part of the halo can be obtained, in order to give a
possible explanation of the observed rotational properties of the
Milky Way stellar halo.  Our results show that, for satellites
  more massive than $\sim 1/40$ of the main halo, the dynamical
friction has a fundamental role in assembling the final velocity
distributions resulting from different orbits and that retrograde
satellites moving on low inclination orbits deposit more stars in the
outer halo regions end therefore can produce the counter-rotating
behavior observed in the outer Milky Way halo.
\end{abstract}





\section{Introduction}

The Galactic halo has long been considered a single component.  However,
evidences in the past few decades have indicated that it may be more complex
\citep{Pre91,Maj92,Kin94,Car96,Chiba00,Kin07,Miceli08}.
Recently \cite{Carollo07}
confirmed the existence of a two-components halo, analyzing a large sample of
calibration stars from the Sloan Digital Sky Survey (SDSS) DR5 \citep{Ade07}.
According to Carollo et al. the Galactic halo consists of two overlapping
structural components, an inner and an outer halo.  These components exhibits
different spatial density profiles, stellar orbits and stellar metalicities.
In particular the inner-halo stars show a small (or zero) net prograde
rotation around the center of the Galaxy.  Outer halo stars possess a clear
retrograde net rotation.

The theory of galaxy formation in a Lambda cold dark matter (LCDM)
Universe predicts galactic stellar halos to be built from multiple
accretion events starting from the first structures to collapse
\citep{White78,Searle78}.  Halos, composed of both dark and baryonic
matter, grow by merging with other halos. While the gas from mergers
and accretions loses its energy through cooling and settles into a
disk, the non dissipative material (accreted stars and dark matter)
form a halo around the Galaxy.  In a LCDM universe, a dark matter halo
big enough to host the Milky Way contains 300+/-100 subhalos
\citep{Diemand04} \citep[such a large number of subhalos is related to
  the so-called "missing satellites problem",][]{Moore99}.

There is significant evidence of past accretions into the Milky Way
\citep{Yanny03,Belo06,Grill09}, and therefore we will accept the
scenario that the Milky Way halo has been formed by subhalos accretion
and we focus on possible origin of a retrograde (counter-rotating)
outer stellar halo, defined as the set of stars orbiting around the
Galaxy at large distance from the disk plane, say $z \ga 15$ kpc:
\citet{Carollo07} found that this distance separates the inner and the
outer halo. It is expected that such halo is made of stars stripped
from merging or disrupted satellites of low mass (\citet{Zolotov09};
for a detailed study of accreting events onto a Milky-Way like halo,
see also \citet{BoyKol10}). However, from cosmological simulation,
prograde and retrograde mergers are approximately equally likely
\citep{Sales07,Read08}.

In order to determine if, and in what condition, we
can obtain a retrograde signal in the stellar distribution in the outer halo,
we simulate minor mergers of a satellite halo onto a main halo of Galactic
mass, with a mass ratio $M_{\rm primary}/M_{\rm satellite}
\sim 40$. We put
the satellite on two orbits, one prograde and one retrograde. After the
satellite completes its merging/disruption process, we identify stars at
distance larger than 15 kpc from the disk plane and simply analyze their
rotation velocities.  Our main finding is that a counter-rotating stellar halo
naturally arises when minor mergers happen on orbits with a low inclination
with respect to the disk plane.

The plan of the Letter is the following. In Section \ref{sec:sim}, we describe
our simulations; in Section \ref{sec:res}, we give our results on
counter-rotating outer halo stars, and in Section \ref{sec:concl} we draw our
conclusions.

\section{Simulations}
\label{sec:sim}

We use a primary Dark Matter (DM) halo containing a stellar, rotating
exponential disk. The DM halo has a NFW \citep{NFW97} radial density profile,
and a mass, radius and concentration appropriate for a Milky Way like DM halo
at redshift $z=0$.  DM particles have velocities given by the local
equilibrium approximation \citep{Hernquist93}. Into our halo, we embed a
truncated stellar disk, having an exponential surface density law: \( \rho
_{\rm stars}=\rho _{\rm 0}\exp -(r/r_{\rm 0}) \) where \( r_{\rm 0} \) is the
disk scale length, and \( \rho _{\rm 0} \) is the surface central density.
We obtain each disk particle's position using the rejection method by
\cite{Press86}; the disk is in gravitational equilibrium with the DM halo (see
\cite{Curir99} for further details).
We choose for the minor merger satellite a mass ratio of
$\approx 40$, similar to the estimated mass ratio of the LMC to the Milky Way
halo. The satellite contains a stellar bulge, with a Hernquist radial
density profile. We realized our DM+bulge satellite configuration as in
\cite{Alvaro08}. All the physical parameters of our merger are listed in
Table~\ref{table:ic}.

\begin{table*}
  \caption{ Properties of main and satellite halos. Column 1: Halo. Column 2:
    Virial mass, in $M_\odot$; here, we refer all our virial quantities to an
    overdensity of 200 times the mean cosmic density. Column 3: Virial radius,
    in kpc. Column 4: NFW concentration parameter. Column 5: Truncation radius
    of the main halo; the secondary halo has an exponential cut-off in
    density. Column 6: Disk scale radius (main halo only). Column 7: Disk
    truncation  radius, in kpc. Column 8: Disk
    stellar mass, in $M_\odot$. Column 9: Hernquist scale radius, in
    kpc. Column 10: Mass of the stellar bulge, in $M_\odot$.  }
\begin{tabular}{c c c c c c c c c c}
\hline\hline Halo & $M_{\rm 200}$ & $R_{\rm 200}$& $C_{\rm 200}$ & $R_{\rm trunc}$ & $r_0$ & $r_{\rm disk}$ & $M_*$ & $a_b$ & $M_b$\\
\hline
Main & $10^{12}$ & 165 & 7.5 & 1300 & 4 & 20 & $5.7 \cdot 10^{10}$ & - & - \\
Satellite & $2.4 \cdot 10^{10}$ & 47 & 8.54 & - & - & - & - & 0.709 & $2.4 \cdot 10^9$ \\
\hline
\end{tabular}
\label{table:ic}
\end{table*}

We simulated prograde mergers, in which a satellite co-rotates
with respect to the disk spin, and retrograde ones with
a counter-rotating satellite.  We
chose two orbits used in \cite{Read08} for studying the thickening of the disk
due to the same kind of minor merger: a low-inclination one, with a 10 degree
angle with the disk plane, and a high inclination one with a 60 degree angle.
Initially, the center of the primary halo stays in the origin of our
coordinate system and the satellite is in (x,y,z) = (80.0, 0.27, -15.2) kpc
for the low-inclination orbit and (15.0, 0.12, -26.0) kpc for the
high-inclination one. The (x,y,z) components of the velocity of the satellite
are, in the prograde case, ( 6.3, -62.5, 0.35) km/s for the low-inclination
orbit and (-1.2, 80.1, 2.0) km/s for the high-inclination one. The retrograde
orbits have the y-component of the velocity inverted.

The
secondary has always a spin parameter $\lambda=0$, where we define
$\lambda=J/ (\surd{2}MVR)$, with $M$ being the mass inside a radius $R$ and
$V=GM/R$ the circular velocity, as in \cite{Bullock01}. We use simulations
with a primary halo with a spin parameter $\lambda=1$ (simulations ``A'') and
$\lambda=0$ (simulations ``B'').  We assign the angular momentum to DM
particles using a rigid body rotation profile. The angular momentum of DM
particles is always aligned with that of the stellar disk.

Our primary halo has $10^6$ DM particles inside the virial radius ($\sim
2.5\cdot 10^6$ in total) and one million star particles in the exponential disk,
with mass $M_{\rm DM}=10^6$ M$_\odot$ and $M_*=5.97 \times10^4$ M$_\odot$
respectively.  Our DM+Bulge secondary halo has $1.1 \cdot 10^5$ DM
particles and $10^5$ bulge star particles, with masses $M_{\rm sat}=1.95
\times 10^5$ M$_\odot$ and $M_{\rm bulge}=2.38 \times 10^4$ M$_\odot$.

We use a Plummer-equivalent gravitational softening length $\varepsilon= 0.5$
kpc, and $\varepsilon=0.25$ kpc for bulge star particles.
We run all our simulation using the public parallel
Treecode GADGET2 \citep{Springel05}.

%In our full suite of numerical experiments, we have DM-only simulation at
%lower resolution (10 times less particles in each component) and at higher
%resolution (5 times more particles). 
To test the convergence of our results with resolution, we re-run our
set ``A'' with ten times more particles in the satellite halo.  We
also changed the spin parameter of primary and secondary halos, its
radial distribution using the profile from Bullock \citep{Bullock01},
and their coupling.

We analyze positions and kinematic of the bulge stars, once the
satellite completes it merging with the primary halo \footnote{We
  define a merger to be complete when the z coordinate of the CM of
  all satellite stars remains within 2 kpc from the initial disk
  plane.}. We run all our simulations for $t=4.63$ Gyrs, corresponding
to $\sim 16$ dynamical timescales of the main halo.

\section{Results}
\label{sec:res}

In all cases, in our simulations the satellite is slowed down by dynamical
friction exerted on it by both disk and halo particles. At the first
pericenter, tidal forces deform the satellite, redistributing its particle's
energy (violent relaxation), and strips away some stellar particles (tidal
stripping). The process continues at each subsequent passage to the
pericenter, until no recognizable self-gravitating structure is present
anymore. Stripped star particles tracks the orbital pattern of the
satellite. The spatial distribution of stars depends upon the dynamical
history of the satellite to which they belonged, i.e., how quickly it loses
its orbital energy, how strongly it gets disrupted by tidal forces, and how
these effects modify the orbit itself during the merger process.

We rotate our coordinate system so that the stellar disk lies in the X-Y
  plane. The origin is in the disk center of mass.
To detect a rotation signal in the outer halo stars, we simply
took all the star particles initially belonging to our satellite, and having a
coordinate $Z>15$ kpc or $Z<-15$ kpc. We then calculate the rotation velocity
of such particles in the disk plane.

%\footnote{In some runs, the disk tilts with
%  respect to its initial configuration. We always use the simulation
%  coordinate system in this work, i.e. rotation is referred to the initial
%  disk plane. This is to avoid that the calculated $v_\phi$ of the {\it same}
%  particle, at different times, spuriously changes due to the change of the
%  coordinate system. This however happens only in one high-inclination set of
%  simulations}

\subsection{Distributions of $v_\phi$}
\begin{figure}
\epsscale{1.4}
\plotone{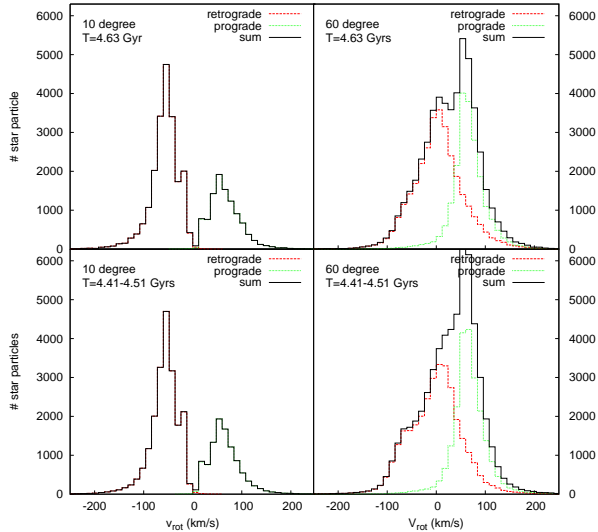}
\caption{ Histograms of rotation velocities for star particles in the outer
  halo, at the end of the set A of merger simulations. Upper panels show the
  histograms at the final time, lower panel show the same histograms, averaged
  over five simulation snapshots. In red (dashed line), we plot the histogram
  obtained for our retrograde orbits; in green (dotted line), those for our
  prograde orbits. In black (solid line) we show the sum of the two. Left
  column is for the low inclination orbits, right column for the high
  inclination ones. We used 50 velocity class, equispaced, between -300 and
  300 km/s in all histograms.
\label{fig:spin1}
}
\end{figure}

In Figure \ref{fig:spin1}, we show histograms of the rotation velocity
obtained in the four simulations of the set A: the satellite is either
co-rotating or counter-rotating with respect to the disk, and the orbit has
either a low or a high inclination
with respect to the disk plane. In the upper panels, we show histograms at the
final time of our simulations; in the lower ones, we performed an average over
five consecutive snapshots, ($\sim 100$ Myr) to get
rid of possible sampling effect on the particle orbits which can rise using a
particular instant of time. We show both the distribution of stars from single
simulations, and the one resulting from taking together star particles from
prograde and retrograde orbits having the same initial inclination. The latter
gives us a hint on the possible rotation signal origin in the case a similar
number of prograde and retrograde minor accretion events happen during the
formation history of a galaxy, under the hypothesis that their inclination is
the same.
It is immediately clear from the Figure that our couple of low inclination
orbits show an excess of counter-rotating stars in the outer halo.

Table \ref{tab:npart} reports the number of satellite star particles in the
positive and negative peak in all of our cases, and the number of particles
having rotation velocity $v_{\rm rot}< -10$ km/s and $v_{\rm rot}>10$ km/s. It
also reports the same numbers obtained from our set B, in which the DM halo
has no spin. We give such numbers for the final time of our simulations. The
counter-rotating excess signal, defined as the fraction of the number of
counter- to co- rotating satellite stellar particles, is 3.39 at peaks and 1.98
overall for the low-inclination orbits, while it is 1.3 (at peaks) for the
high-inclination ones.  Note that, for high-inclination orbits, the
  retrograde case shows a peak at a rotation velocity near to $v_{\rm
    rot}=0$. This is because high inclination orbits have a larger impact
parameters. The disk responds by tilting more than in the
low-inclination case, and as a result,  rotation velocities in the disk
reference frame are shifted towards more positive values. This effect is also
present in our low-inclination runs, but it is much smaller because the tilt
of the disk is smaller.

\begin{table*}
\caption{ 
Number of outer halo satellite star particles in the positive and negative
peak in all of our cases, and the number of outer halo satellite star
particles having rotation velocity $v_{\rm rot}< -10$ km/s and $v_{\rm
  rot}>10$ km/s in our low inclination run. 
Results are for the final time of our simulations. For each
inclination, we considered together prograde and retrograde mergers. First
row: numbers of satellite star particles having rotation velocities smaller
than $-10$ km/s or higher than $10$ km/s, for the low inclination case. Second
row: number of star particle in the positive and negative peak, for the low
inclination case. 
Third row: number of star particles in the positive and
negative peak, for the high inclination case. Second column: number of
counter-rotating satellite stars when the DM halo has a spin
$\lambda=1$. Third column: number of co-rotating satellite stars when the DM
halo has a spin $\lambda=1$.  Fourth column: number of counter-rotating
satellite stars when the DM halo has no spin, $\lambda=0$. Fifth column:
number of co-rotating satellite stars in the same $\lambda=0$ cases.
}
\begin{tabular}{c c c c c}
\hline\hline Distribution & $\lambda=1$ counter & $\lambda=1$
co & $\lambda=0$ counter& $\lambda=0$ co\\
\hline
10deg TOTAL $v_*>|10|$ km/s & 20292 & 10237 & 16670 & 13033 \\
10deg PEAK & 6510 (-54) & 1917 (+54) & 4455 (-54) & 3419 (54) \\
60deg PEAK & 3581 (+6) & 4010 (64) & 2741 (-6) & 3071 (78) \\
\hline
\end{tabular}
\label{tab:npart}
\end{table*}

It remains to be determined if the counter-rotating signal is caused by the
interaction with the stellar disk of the main halo, or by the DM of the halo
itself.

\begin{figure}
\epsscale{1.4}
\plotone{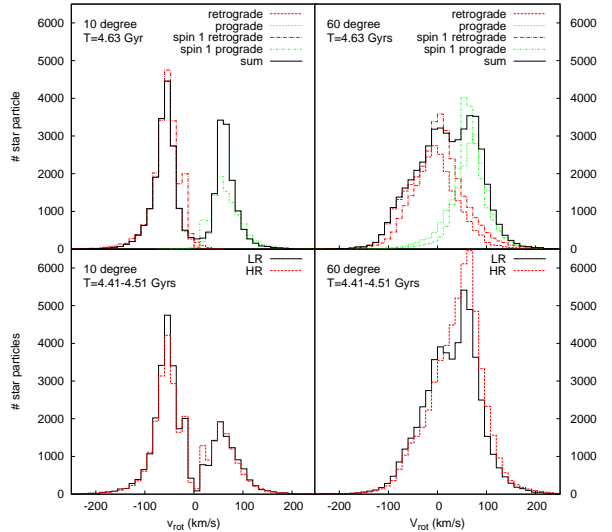}
\caption{ 
Upper row: histograms of rotation velocities for star particles in
the outer halo, at the end of the set B of merger simulations, in which the
main halo has no spin.  In red (dashed line), we plot the histogram obtained
for our retrograde orbits; in green (dotted line), those for our prograde
orbits. In black (solid line) we show the sum of the two. Left panel is for
the low inclination orbits, right panel for the high inclination ones. We
also show the histogram for the case of spin 1 (dashed-dotted lines).
Lower row: histogram of rotation velocities for star particles in the outer
halo, for our set A, at the basic resolution (``LR'', black continuous line)
and at a ten times better mass resolution (``HR'', red dashed line).
\label{fig:spin0}
}
\end{figure}

The upper row of Figure \ref{fig:spin0} shows the results of the same analysis
performed on our set A, but in the case in which no halo spin is present (set
B). Here we only show the results for our last snapshot; as in Figure
\ref{fig:spin1}, averaging over five snapshots makes no appreciable
difference.  Also in our set B, low inclination retrograde orbits produce more
counter-rotating star particle than co-rotating stars produced by prograde
orbits. Again, this is not the case for high inclination orbits. From Figure
\ref{fig:spin0}, it is clear that the excess of counter-rotating star is 
  clearly smaller: and this is due to the fact that {\it more} co-rotating
stars are produced by prograde orbit in our no spin case than in spin 1 case.
From Table \ref{tab:npart}, the counter-rotating excess signal is 1.60 at
peaks and 1.59 overall in the low inclination case, and 1.17 at peaks and 1.02
overall in the high inclination one.

Therefore, both disk rotation and halo spin contribute to the slowing-down of
prograde orbits and to the consequent smaller amount of high-energy star
particles stripped from satellites that can reach the outer halo. 
%However, disk
%rotations appear to be the main driver of such an effect.

The lower row of Figure~\ref{fig:spin0} shows that our result does not depend on
the mass resolution of our satellite halo. Even using 10 times more particles
in the secondary, only our low inclination couple of mergers shows an excess
of counter-rotating stars in the outer halo.

\subsection{Description of the mergers}
Two effects are acting on the satellite halo in these kind of mergers:
dynamical friction and tidal disruption. The first one is exerted both by the
main halo DM particles and by the disk star particles. The second is most
important near the center of the main halo, where the gravitational potential
is stronger. But, dynamical friction depends upon the details of the satellite
orbits. It is already known
\citep{Quinn86,Walker96,Huang97,Velazquez99,Alvaro08} that prograde orbits
tends to decay faster than retrograde ones. The dynamical friction {\it force}
goes as $F_{\rm dyn} \propto 1/v_s^2$ \citep{BinneyTremaine}, where 
$v_s$ is the velocity of the satellite relative to
the field particles; 
retrograde satellites have higher
$v_s$ with respect to prograde one, since in the first case the rotation velocity of the
satellite is opposite to that of the disk and the main DM halo particles.
Particles stripped from the satellites will remain on the orbit on
which the satellite was when they were stripped. Therefore, retrograde
satellites ``deposit'' more star particles in the outer halo regions,
producing the signal we observe in our simulations, since their orbits have a
longer decay time. Obviously, to obtain this effect, the {\it orbital}
velocity must lie on the disk plane \footnote{DM spin is parallel to
  the disk spin: also the effect of the DM halo spin is maximum in such a case}, which is almost the case for our low
inclination orbits.  High inclination encounters do not show such a
behavior.

\begin{figure}
\epsscale{1.6}
\plotone{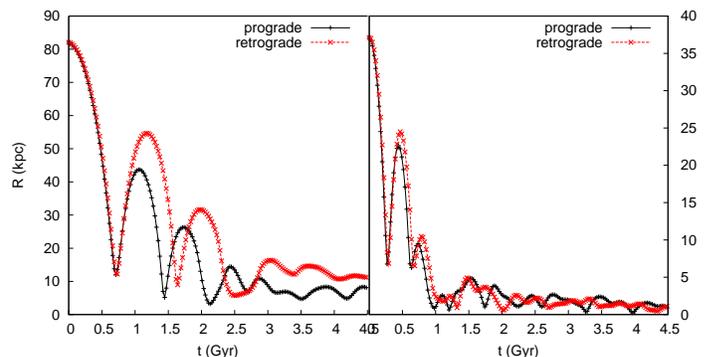}
\caption{
Position of the center of mass of all star particles belonging to the satellite,
as a function of time. Left panel shows the result for our low inclination
orbits, right panel for the high inclination ones. The center of mass position
makes sense only until the satellite is not disrupted; this happens after
$\sim 3$ Gyr in our low-inclination prograde simulation, $\sim 4$ Gyr in 
our low-inclination retrograde one, while in the high inclination cases
cases the satellite is disrupted already after $\sim 1$ Gyr.
\label{fig:cms}
}
\end{figure}

In Figure \ref{fig:cms}, we plot the position of the center of mass of all the
star particles belonging to the satellite as a function of time.  In our low
inclination mergers (left panel), the prograde orbit decades faster: already
after 1.5 Gyr, the difference with the behavior of the retrograde satellite is
clear. In the high inclination cases the effect, though present, is much
smaller, as expected. 

We also verified that the effect we show here does not depend upon the exact
angular momentum distribution of particles in DM halo, by repeating our
experiments using a Bullock (2001) profile instead of a rigid rotation one for
the DM rotation velocity. Finally, the initial distance at which we put
our satellite does not have a great impact on our results: we repeated the
low-inclination mergers, putting the satellite at coordinates (29.5, 0.27,
-5.2) kpc, thus much nearer to the center, and obtained the same qualitative
behavior.

\section{Conclusions}
\label{sec:concl}

We performed controlled numerical simulations of minor merger events. Our aim
was to determine if an excess of counter-rotating stars in the outer stellar
halo can be produced by couples of mergers having orbits with identical
inclination and opposite initial rotation. Our primary halo has an exponential
stellar disk and its dark matter has a NFW radial density profile; our
satellite's DM has an NFW density profile, and contains a stellar spheroid
with a Hernquist density profile. The mass ratio of our merger is $M_{\rm
  primary}/M_{\rm satellite} \sim 40$. We simulate low inclination encounters,
in which the angle of the satellite's orbit with the disk plane is 10 degrees,
and high inclination ones (60 degrees). We use prograde and retrograde orbits
and also vary the DM spin parameter of the main halo. DM spin is aligned with
the disk rotation.  Here we define all stars having a distance $z>15$ kpc from
the disk plane as belonging to the outer stellar halo.

Our main results are the following:
\begin{itemize}

\item low inclination mergers do produce an excess of counter-rotating
  satellite stellar particles in the outer halo, independently on the spin of
  the DM;
\item our 60 degrees mergers with our 1:40 mass ratio are not
  able to give a significant counter-rotating signal, also owing to
  the disk tilt produced by the merger itself;
\item the fraction of counter-rotating to co-rotating satellite stars in the
  outer stellar halo is higher if the DM has spin.

\end{itemize}

From our controlled experiments, we now have an indication of the possible
origin of the counter-rotation of stars observed in the Galactic outer
halo in the context of a hierarchical $\Lambda$CDM model of galaxy formation, in which many
major mergers happen during the history of the Galaxy. 
Even if, statistically, the number of prograde and retrograde minor
merger is the same, still a counter-rotating signal can arise {\it if such
  mergers predominantly happen along low inclination orbits}. Since matter
accretion, in a CDM dominated Universe, mainly occurs along filaments, this
will be the case {\it if the galaxy disk is co-planar} to the (majority of)
filaments. The disk-filament alignment issue is still debated \cite[see e.g.][]{Brunino07}: from our results,
we expect that if the galactic disk were perpendicular to the main accretion
streams, no counter-rotating signal should be observed in the outer halo star
distribution. 

Our orbits are {\it not} cosmologically motivated, since the aim of our
experiment is to determine {\it if} and {\it in what cases} an excess of
counter-rotating stars in the outer halo can be produced.  Of course, in a
realistic case mergers will occur with orbits having a number of different
inclinations, giving rise to a velocity distribution which will not show such
a clear, double peaked signal as that detected in the present work. A detailed
study of the orbital parameters of minor mergers in a statistically
significant set of cosmological Galaxy-sized halo cosmological accretion
histories is needed before a quantitative comparison between theory
expectations and observation can be performed.

\acknowledgments 

We acknowledge useful discussions with Daniela Carollo, Stefano Borgani,
Gabriella De Lucia, Antonaldo Diaferio and Alessandro Spagna The simulations
were carried out at CASPUR, with CPU time assigned with the ``Standard HPC
grant 2009'' call, and at the ``Centro Interuniversitario del Nord-Est per il
Calcolo Elettronico'' (CINECA, Bologna), with CPU time assigned under an
INAF/CINECA grant. A.C. acknowledge the financial support of INAF through the
PRIN 2007 grant n. CRA 1.06.10.04 "The local route to galaxy formation".

%\bibliographystyle{apj}
%\bibliography{master}

\clearpage

\end{document}